\documentclass[aip,jcp,preprint]{revtex4-1}
\usepackage{graphicx}
\usepackage{amsmath}
\usepackage{amssymb}

\begin{document}
\title{Effects of the dielectric discontinuity on the counterion distribution in a colloidal suspension}

\author{Alexandre P. dos Santos}
\affiliation{Instituto de F\'isica, Universidade Federal do Rio Grande do Sul, Caixa Postal 15051, CEP 91501-970, Porto Alegre, RS, Brazil}

\author{Amin Bakhshandeh}
\affiliation{Instituto de F\'isica, Universidade Federal do Rio Grande do Sul, Caixa Postal 15051, CEP 91501-970, Porto Alegre, RS, Brazil}
\affiliation{Department of Physical Chemistry, School of Chemistry, University College of Science, University of  Tehran, Tehran 14155, Iran}

\author{Yan Levin}
\email{levin@if.ufrgs.br}
\affiliation{Instituto de F\'isica, Universidade Federal do Rio Grande do Sul, Caixa Postal 15051, CEP 91501-970, Porto Alegre, RS, Brazil}

\begin{abstract}
We introduce a new method for simulating colloidal suspensions with spherical colloidal particles of dielectric
constant different from the surrounding medium. The method uses exact calculation of the Green function 
to obtain the ion-ion interaction potential in the presence of a dielectric discontinuity 
at the surface of the colloidal particle. 
The new method is orders of magnitude faster than the traditional approaches based on series expansions of 
the interaction potential.
\end{abstract}

\maketitle

\section{Introduction}

Colloidal suspensions are of fundamental interest for various applications.  One of the basic problems of 
colloidal science is how to stabilize a lyophobic colloidal suspension against
flocculation and  precipitation. A common approach is to synthesize 
particles with acidic or basic charged groups on the surface~\cite{MiFo00,MiMa01}.  
When placed in a polar medium such as water, these groups become ionized and the particles acquire a net charge.  Repulsion between
like-charged colloidal particles then prevents them from coming into a close contact where the short-range van der Waals forces
become important.   Addition of electrolyte to colloidal suspension leads
to screening of the Coulomb  repulsion~\cite{Le02}. At critical coagulation concentration~(CCC), the repulsive energy barrier
disappears and the van der Waals forces drive colloidal coagulation and precipitation~\cite{LoJo03,LoSa08,PeOr10,DoLe11}.   It is also well known
that addition of even very small amount of multivalent ions leads to a rapid precipitation.  The correlation 
induced attraction between the colloidal particles produced by the multivalent ions is sufficient to precipitate colloidal suspensions even without taking into account the van der Waals forces. The like-charge
attraction has been extensively explored in colloidal and polyelectrolyte literature~\cite{RoBl96,LiLo99,WuBr99,GeBr00,SoDe01}.  A related phenomenon known as the 
charge reversal has also attracted a lot of attention over the recent years~\cite{LoSa82,PiBa05,DiLe06,DiLe08,GuGo10,DoDi10}.  
In this case electrostatic correlations
result in a strong colloid-counterion association\cite{Le02}.  The counterion condensation can be so significant as to reverse the electrophoretic mobility of colloidal particles~\cite{QuCa02,FeFe05}. 

Most of the theoretical work on stability of colloidal suspensions and the charge reversal, 
however, neglects the effects of the dielectric discontinuity
at the particle/solvent interface.  In fact, in many colloidal suspensions the static dielectric constant of colloidal particles can
be $20$ to $40$ times lower than the static dielectric constant of the surrounding water.  
This means that an ion in the vicinity of colloidal 
surface will encounter a strong ion-image repulsion.  This repulsion can significantly affect the effective charge
of the colloid-counterion complex and thus modify the 
colloid-colloid interaction
potential.  The polarization effects, however, have been mostly
neglected in almost all of the theoretical studies.  The reason for this is  that it is very hard to include the 
dielectric discontinuities in anything but the
simplest planar geometry.  Thus, even to perform a Monte Carlo simulation that accounts for the dielectric 
discontinuity requires a significant computational effort. Some years  ago Linse~\cite{Li86} proposed to account for the
induced charges by treating the low dielectric colloidal particle as if it was an ``inverse'' conductor.  
It is well known that if one places a charge near a conducting sphere, a surface charge will be induced on the sphere~\cite{Ja99}.
The field produced by the surface charge is exactly equivalent to the field produced by two point charges of opposite
sign, one located at the spheres inversion point and  another at its center.  In the case of a conductor, the charge at the
inversion point has the opposite sign to the charge placed outside the sphere, so that this charge is attracted to the
conductor.  Linse suggested that the low dielectric sphere in water 
can be treated as an inverse conductor, meaning that the same construction should apply to locate the images, but that
their sign will be the opposite of the images inside the conducting
sphere.  This, however, is not quite right.   Because of the complicated boundary
conditions (BCs) imposed by the Maxwell equations at the dielectric interface, one can not satisfy the 
BCs with only two pointlike  image charges. In fact one needs an infinite number of images~\cite{Li92,No95}. 

Polarization effects have also been explored in ionic liquids~\cite{ReLa07,ReMa08,LoSk08,OuLa10} and for  polyelectrolyte
adsorption~\cite{Me04,Me06,SePo09}  in a slab geometry. 
In this paper we will derive the explicit 
inter-ionic interaction potential
which very accurately accounts for the dielectric discontinuity for spherical colloidal particles. 
We compare our results with the Monte Carlo simulations of Messina~\cite{Me02}, who obtained the ion-ion 
interaction potential as an infinite series in Legendre polynomials.
In the final part of the paper we will analyze the effect of 1:1 electrolyte on the  distribution of trivalent 
and monovalent counterions near the colloidal surface.

\section{Method}

\begin{figure}[t]
\begin{center}
\includegraphics[width=4.5cm]{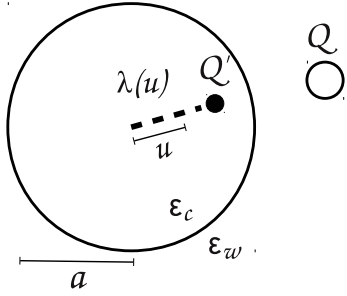}
\end{center}
\caption{Illustrative representation of an ion of charge $Q$ near a colloidal particle and the induced image, $Q'$, and 
counterimage, $\lambda(u)$, charges.}
\label{fig1}
\end{figure}

We will use a primitive model of a colloidal suspension in which colloidal particle is represented by a sphere
of radius  $a$ and the dielectric constant $\epsilon_c$.  Water will be modeled as a uniform dielectric of 
permittivity $\epsilon_w$. The system is at room temperature, so that the Bjerrum length, 
defined as $\lambda_B=q^2/\epsilon_w k_BT$, is $7.14$\AA.
Consider an $\alpha$-valent ion of charge $Q=\alpha q$, where $q$ is the proton charge, 
at position ${\bf r}_i$ from the center of the colloidal particle.  The Maxwell equations
require the continuity of the tangential component of the electric field and the continuity 
of the normal component of the displacement field
across the colloid-water interface.  
It is possible to show~\cite{Li92,No95} that this boundary conditions can be satisfied {\it exactly} by placing
an image charge $Q'=\gamma Q a/r_i$ inside the colloid at the inversion point ${\bf r'}_{i}={\bf r}_i a^2/r_i^2$ and a 
counterimage  line-charge $\lambda(u)$, distributed along the line connecting the center of colloid
with the inversion point  ${\bf r'_i}$,  with line-charge density 
\begin{equation}
\lambda(u)=-\dfrac{Q' (1+\gamma)}{2 r'_i}\left(\dfrac{u}{r'_i}\right)^{\frac{\gamma-1}{2}} \ ,
\end{equation}
where $\gamma=(\epsilon_w-\epsilon_c)/(\epsilon_w+\epsilon_c)$ and $0 \le u \le r'_i$ is the distance along the line, see Fig.~\ref{fig1}. Note that this construction does not change the net charge of the colloidal particle, that is the total counterimage charge 
is $-Q'$.        

The electrostatic potential produced by the image charge at an arbitrary position ${\bf r}$ {\it outside colloid} is 
\begin{equation}
\psi_{im}({\bf r};{\bf r}_i)= \dfrac{\gamma \alpha q a  }{\epsilon_w r_i|{\bf r}-\frac{a^2}{r_i^2}{\bf r_i}|} \ ,
\label{image}
\end{equation}
and the electrostatic potential produced by the counterimage line-charge  is
\begin{equation}
\psi_{ci}({\bf r};{\bf r}_i)=\dfrac{a^2}{\epsilon_w r_i}\int_{0}^{1}d \eta \dfrac{\lambda(\eta \frac{a^2}{r_i})}{|{\bf r}-
{\eta \frac{a^2}{r_i^2}{\bf r_i}}|} \ .
\label{exact}
\end{equation}
For ${\bf r}={\bf r}_i$, the integral can be performed exactly in terms of the hypergeometric function  $_2F_1$.  We find the counterimage-ion interaction potential to be
\begin{equation}
\psi_{ci}^{self}({\bf r}_i)=-\dfrac{\gamma \alpha q a}{\epsilon_w r_i^2} \,_2F_1\left( \frac{1}{2}+\frac{1}{2}\gamma , 1 , \frac{3}{2}+\frac{1}{2}\gamma , \frac{a^2}{r_i^2}\right) \ .
\label{exact_self}
\end{equation}

Although  Eqs.~\ref{exact} and \ref{exact_self}
are  exact, they are not very useful for Monte-Carlo or molecular dynamics simulations --- the integral
in Eq.~\ref{exact} must be done numerically for each new configuration of ions,  making  simulations very slow. 
However, we can consider a simplifying approximation.  We note that the dielectric constant 
of a colloidal particle is much smaller than the dielectric constant of the surrounding medium.  
Thus, to leading order in $\epsilon_c/\epsilon_w$ we can take $\gamma \approx 1$.  
In this case the counterimage charge is uniformly distributed, 
$\bar \lambda(u)=-Q'/r'_i$, and the integral in Eq.~\ref{exact} can be performed exactly, yielding the counterimage potential at
an arbitrary position ${\bf r}$,
\begin{equation}
\bar \psi_{ci}({\bf r};{\bf r}_i)=\dfrac{\alpha q}{\epsilon_w a}
\log \left(\frac{r r_i -{\bf r} \cdot {\bf r_i}}{a^2-{\bf r} \cdot {\bf r_i}+\sqrt{a^4-2 a^2 ({\bf r} \cdot {\bf r_i})+r^2 r_i^2}}\right) \ ,
\label{app}
\end{equation}
where the over-bar is used to denote the uniform line-charge approximation.  
The ion-counterimage interaction potential also reduces to a simple equation,
\begin{equation}
\bar \psi_{ci}^{self}({\bf r}_i)=\dfrac{\alpha q}{\epsilon_w a} \log \left( 1-\frac{a^2}{r_i^2} \right) \ .
\label{app_self}
\end{equation}

\section{Monte Carlo Simulations}

The simulations are performed inside a spherical Wigner-Seitz~(WS) cell of radius $R$ with a colloidal particle of charge  $-Zq$ placed at the center. The cell also contains  $N=Z/\alpha$ $\alpha$-valent counterions each of diameter $d$.  The electrostatic potential produced at position ${\bf r} $ by an ion located at ${\bf r}_i$ is
\begin{equation}
\phi({\bf r};{\bf r}_i)= \dfrac{\alpha q}{\epsilon_w |{\bf r}-{\bf r}_i|} + \dfrac{\gamma \alpha q a  }{\epsilon_w r_i|{\bf r}-\frac{a^2}{r_i^2}{\bf r}_i|} + \gamma \bar \psi_{ci}({\bf r};{\bf r}_i) \ ,
\label{phi}
\end{equation}
where the first term is the electrostatic potential produced by the ion and the second and the third terms are the potentials 
produced by the image and the counterimage charges, respectively.  The first two terms of Eq.~\ref{phi} are exact.  In the third term we have used the condition of charge neutrality
to correct the ion-counterimage interaction from Eq.~\ref{app} by including a prefactor $\gamma$ in 
front of $\bar \psi_{ci}({\bf r};{\bf r}_i)$.  This, then, is
the Green function for the present geometry. The interaction potential between two ions  
$i$ and $j$ is  $\alpha q \phi({\bf r}_i;{\bf r}_j)$. The work required to bring all the ions from infinity to their respective positions inside the cell is,
\begin{equation}
U=\sum_{i=1}^{N}-\dfrac{Z\alpha q^2}{\epsilon_w r_i} + \sum_{i=1}^{N}U_{i}^{self} + \sum_{i=1}^{N-1}\sum_{j=i+1}^{N} \alpha q\phi({\bf r}_i ;{\bf r}_j) \ ,
\label{total}
\end{equation}
\begin{equation}
U_{i}^{self}=\dfrac{\gamma \alpha^2 q^2 a  }{2\epsilon_w (r_i^2-a^2)} +  \dfrac{\alpha q \gamma\bar \psi_{ci}^{self}({\bf r}_i)}{2} \ ,
\end{equation}
where $U_{i}^{self}$ is the interaction energy of the ion $i$ with its image and counterimage charges.

\begin{figure}[t]
\begin{center}
\includegraphics[width=6.5cm]{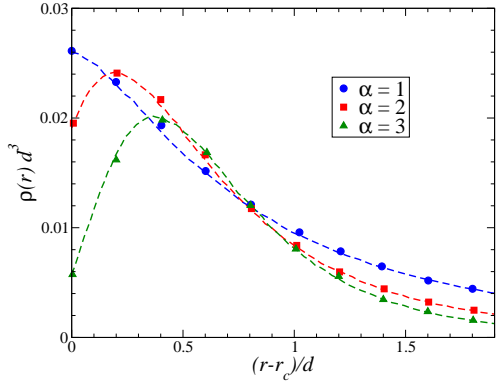}
\end{center}
\caption{The dashed lines represent the density profiles obtained using the present method and the symbols represent the profiles obtained by Messina~\cite{Me02}. The parameters of the simulations are: $\epsilon_w=80$, $\epsilon_c=2$, $d=3.57$\AA, $a=7.5d$, $R=40d$, $Z=60$, and $r_c=a+d/2$ is the contact distance.}
\label{fig2}
\end{figure}
We use the Eq.~\ref{total} in a typical Metropolis algorithm~\cite{Allen}, with $10^5$ MC steps for equilibration 
and $10^4$ steps for production. We obtain the ionic density profiles dividing the WS cell in 
volumetric bins and counting the average number of particles in each bin
for all uncorrelated configurations. In Fig.~\ref{fig2} the profiles 
are compared with the ones obtained by Messina~\cite{Me02}. 
The agreement between the two simulations is excellent, with a huge gain in computational time.
To take into account the correct boundary conditions, Messina calculated the inter-ionic interaction potential as 
an infinite series in Legendre polynomials.  
This method is extremely slow, since one needs to calculate hundreds of terms of the infinite
series for each new configuration in order to obtain a good convergence.  To compare the time of processing, we perform the simulation of Messina for  parameters described in Fig.~\ref{fig2} with trivalent ions. Even for a very small number of counterions --- twenty counterions --- used by Messina, expansion in Legendre polynomials is $\approx 458\times$ slower than the method presented in the current paper.
To speed up the simulations Messina tabulated the counterion-counterion interaction potential.  
Nevertheless, his approach remains at least one order of magnitude slower than our Green function method, and is significantly more difficult to extend to larger system sizes.  

\begin{figure}[t]
\begin{center}
\includegraphics[width=6.5cm]{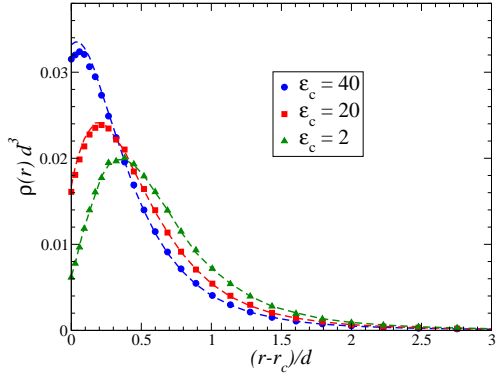}
\end{center}
\caption{Density profiles for various $\epsilon_c$. The symbols represent the density profiles obtained using 
the exact counterimage line-charge distribution, while the dashed lines are calculated using the approximate
uniform counterimage line-charge distribution, Eqs.~\ref{app} and \ref{app_self}.  
The parameters of the simulations are the same as in Fig.~\ref{fig2}, for $\alpha=3$.}
\label{fig3}
\end{figure}

The approximation of the counterimage charge by a uniform line-charge density, 
should work very well for colloids of low dielectric constants, 
which are of most practical interest. However, it is interesting to examine up to what value of $\epsilon_c$ does
this approximation remains accurate.   Using the exact numerical evaluation of the integral in Eq.~\ref{exact}, 
and the hypergeometric representation
of the counterimage-ion interaction potential Eq.~\ref{exact_self}, we have  
performed the simulations for different values of $\epsilon_c$ using the exact numerically calculated interaction potential
and compared the counterion density profiles with the ones obtained in simulations with the approximate interaction 
potentials, Eqs.~\ref{app} and \ref{app_self}. In Fig.~\ref{fig3} we show the results of these simulations. 
We see that the approximation works very well up to  $\epsilon_c \approx 20$, which are in the range of 
most practical interest.
\begin{figure}[b]
\begin{center}
\includegraphics[width=6.5cm]{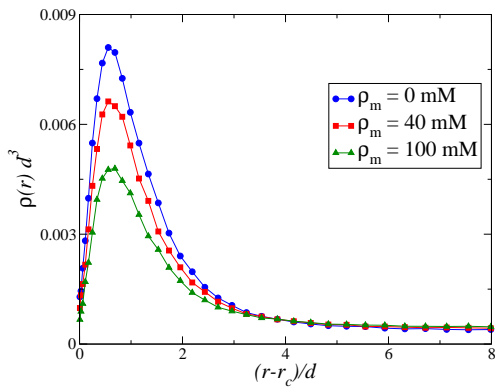}
\end{center}
\caption{Density profiles of trivalent counterions for various concentrations of 1:1 electrolyte. 
The parameters of the simulations are the same as in Fig.~\ref{fig2} for $R=25d$, $\alpha=1$ and $Z=40$. 
The 3:1 concentration is $20$~mM.}
\label{fig4}
\end{figure}

The simulations using the  method developed in the present work are so quick that it is easy to study systems which contain
mixtures of multivalent and monovalent electrolytes.  We next consider a WS cell that
contain 3:1 electrolyte at concentration  $\rho_t$ and 1:1 electrolyte at concentration $\rho_m$. The number of trivalent counterions inside the system is $N_t=\rho_t \frac{4\pi}{3}(R^3-a^3)$, the number of monovalent counterions is $N_m=\rho_m \frac{4\pi}{3}(R^3-a^3)$ and the number of monovalent coions is $N_-=3N_t+N_m$. 
In Fig.~\ref{fig4}, the density profiles of 3:1 salt cations are presented for various 
concentration of 1:1 salt. As expected, with increase of the monovalent salt concentration, more trivalent cations prefer to be solvated in the bulk of suspension, where their electrostatic self-energy is screened most effectively by the
other ions~\cite{PiBa05,DoDi10}.
\begin{figure}[h]
\begin{center}
\includegraphics[width=6.5cm]{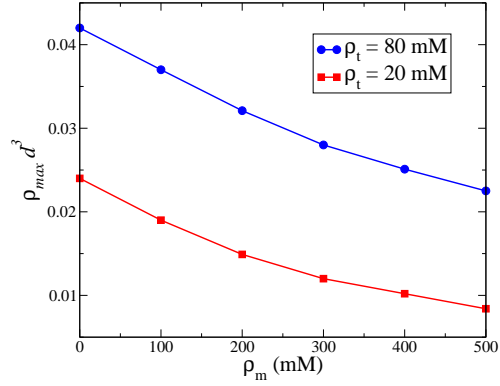}
\end{center}
\caption{Maximum density of trivalent salt counterions as function of concentration of 1:1 electrolyte. 
The parameters of the simulations are: $\epsilon_w=80$, $\epsilon_c=0$, $d=4$\AA, $a=30$\AA, $R=70$\AA\ and $Z=90$.}
\label{fig5}
\end{figure}
\begin{figure}[b]
\begin{center}
\includegraphics[width=6.5cm]{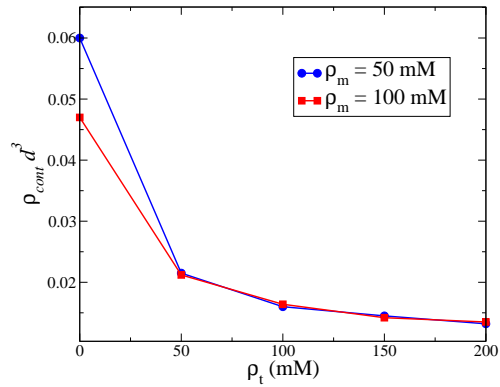}
\end{center}
\caption{The density at contact of monovalent counterions for varying concentrations of 3:1 electrolyte. 
The parameters of the simulations are the same as in Fig.~\ref{fig5}.}
\label{fig6}
\end{figure}

The charge-image repulsion results in density profiles of trivalent ions which have a characteristic
maximum near the colloidal surface. In Fig.~\ref{fig5}, we examine the effects 
of 3:1 and 1:1 electrolyte on the maximum density of trivalent counterions near the colloidal surface.  Again we
see that increasing the concentration of 1:1 electrolyte diminishes the counterion condensation --- resulting in a smaller
counterion density in the vicinity of the colloidal surface.  More surprising, perhaps, is the behavior of
the contact density of the monovalent counterions, Fig.~\ref{fig6}.  We see that at small concentrations of 3:1 electrolyte
the contact density varies significantly with the concentration of 1:1 electrolyte.  This dependence, however, 
rapidly saturates, so that for $50$~mM of 3:1 electrolyte, we no longer see any variation of the contact density
with the concentration of 1:1 salt. The Fig.~\ref{fig6} shows that with increasing 3:1 concentration
the condensed monovalent counterions are rapidly replaced by the trivalent ones.

\section{Conclusions}
We have presented a very efficient method for simulating colloidal suspension composed of lyophobic colloidal particles  
of low dielectric
constant. The method  relies on the exact calculation of the Green function for the spherical geometry. 
The results are in excellent agreement with the earlier simulations of Messina~\cite{Me02}  ---  who used expansion in Legendre polynomials to account for the
dielectric discontinuity at the colloidal surface ---  with a huge gain in the computation time. At the moment, we have only 
implemented the simulation inside  a WS cell geometry.  In the future, an effort should be made to extend the theory to take
into account periodic boundary conditions through the use of Ewald summation.

This work was partially supported by the CNPq, Fapergs, INCT-FCx, and by the US-AFOSR under the grant FA9550-09-1-0283.
\bibliography{ref.bib}

\end{document}